\documentclass[1p]{elsarticle}

\usepackage{hyperref}
\usepackage{amsmath,amssymb, amsfonts,empheq,mathtools,mathrsfs,nccmath}
\usepackage{float,subfig, wrapfig}  
\usepackage{arydshln,multirow, multicol, adjustbox}  
\usepackage{relsize} 
\usepackage{color,xcolor} 
\usepackage[linesnumbered,boxed,ruled,commentsnumbered]{algorithm2e}

\biboptions{authoryear}

\begin{document}

\begin{frontmatter}

\title{An Integrated Early Warning System for Stock Market Turbulence}

\author[mymainaddress,mysecondaryaddress]{Peiwan Wang}
\author[mymainaddress,mysecondaryaddress]{Lu Zong\corref{mycorrespondingauthor}}
\cortext[mycorrespondingauthor]{Corresponding author}
\ead{Lu.Zong@xjtlu.edu.cn}
\author[mymainaddress,mysecondaryaddress]{Ye Ma}

\address[mymainaddress]{Department of Mathematical Sciences, Xi'an Jiaotong-Liverpool University.}
\address[mysecondaryaddress]{111 Ren'ai Road, Dushu Lake Science and Education Innovation District, Suzhou Industrial Park, Suzhou, Jiangsu Province, P.R. China, 215123.}

\begin{abstract}
This study constructs an integrated early warning system (EWS) that identifies and predicts stock market turbulence. Based on switching ARCH (SWARCH) filtering probabilities of the high volatility regime, the proposed EWS first classifies stock market crises according to an indicator function with thresholds dynamically selected by the two-peak method. An hybrid algorithm is then developed in the framework of a long short-term memory (LSTM) network to make daily predictions that alert turmoils. In the empirical evaluation based on ten-year Chinese stock data, the proposed EWS yields satisfying results with the test-set accuracy of $96.6\%$ and on average $2.4$ days of forewarned period. The model's stability and practical value in the real-time decision-making are also proven by the cross-validation and back-testing.
\end{abstract}


\end{frontmatter}

\section{Introduction}\label{sec:introduction}

Due to the Subprime Mortgage crisis, the Shanghai Stock Exchange Composite (SSEC) index experienced one of its greatest falls in the end of 2007. In mid-2015, another Chinese stock market bubble crashed and led to extreme turbulence and instability in the domestic financial environment. As the lasting effect of stock market crises is recognized as the cause of critical society stress and results in increasing financial loads of the government, a systematic model that monitors the economic scenarios of financial markets, and generates early warning signals for potential extreme risks is in heavy demand.

Financial early warning systems (EWSs) are designed to forecast crises via studying pre-turmoil patterns, thus to allow market participants to take early actions to hedge against vital risks. In practice, the target of early warning ranges from individual financial markets, such as the banking sector, the currency and stock markets, to the entire economic system. The modeling of crises are then commonly formulated as a classification problem based on the identified crisis indicators. To design an effective and reliable EWS with true warnings and limited false alarms, two matters need to be delicately addressed, that is the identification of crises and the mechanism of prediction. 

In the previous studies, an EWS is primarily constructed by identifying crises on the basis of either expert opinions or an indicator function describing the market crash. The former approach is widely used in the early studies of EWS, especially those concerning banking and debt crises \citep{klr, kaminsky2006currency, reinhart2011financial, reinhart2013banking, caprio2002episodes, valencia2008systemic, laeven2010resolution, laeven2012systemic, detragiache2001crises, yeyati2011elusive}. Despite that the expert-defined crises are considered to be reliable for long-term predictions \citep{oh2006early}, this paradigm fails to offer an efficient modeling solution as the frequency of observation increases. On the other hand, indicator functions based on a pre-specified threshold are more frequently used to define currency or stock market crashes. \cite{reinhart2011financial} define a currency crisis as the excessive exchange rate depreciation exceeds the threshold value of $15\%$. Alternatively, \cite{eichengreen1995exchange} propose to use the Financial Pressure Index (FPI) to measure the gross foreign exchange reserves of the Central Bank and the repo rate \citep{sevim2014developing}. Currency crises are thus identified as the FPI raises more than $1.5$ \citep{kibritcioglu1999leading}, $2$ \citep{eichengreen1995exchange, bussiere2006towards}, $2.5$  \citep{edison2003indicators} or, $3$ \citep{klr, berg1999predicting, duan2008china}) standard deviations from its long-term mean. In the context of stock EWS, market crashes are indicated by the CMAX index falling below its mean by $2$ \citep{coudert2008does}, $2.5$ \citep{li2015toward}, or $3$ \citep{fu2019predicting} standard deviations. In terms of expressing crises as indicator functions, two major drawbacks emerge in the practical aspect. Despite that the paradigm of handling crises as crashes captures the associated acute loss, it fails to consider the extreme risk that comes along with the volatility jump. Moreover, the selection of crisis thresholds should be handled more delicately taking into account the trade-off between missing crises and false alarms resulted from over-/under-estimated thresholds \citep{babecky2014banking}. 

In terms of the predictive model, three types of methods are commonly applied to generate early warning signals for currency, banking and debt crises, namely the logit-probit regression \citep{frankel, eichengreen, demirg, bussiere2006towards, beckman} , the signaling approach \citep{klr, kaimingrac, berg1999predicting, davis} and machine learning-based models \citep{nag, kim2004korean, celik, yu2010multiscale, giovanis2012study, sevim2014developing}. Among the limited studies on stock markets \citep{fu2019predicting}, \cite{coudert2008does} use logit and multi-logit models to predict stock and currency crises and find the leading effect of risk aversion indicators for stock early warning. \cite{li2015toward} shows the significance of S\&P 500 futures and options in predicting stock crashes basing on a logit model. By combining the logit model and Ensemble Empirical Mode Decomposition, \citep{fu2019predicting} recently develop an EWS for daily stock crashes using investor sentiment indicators and achieve good in-sample and test-set results. Due to the non-linear nature of financial data, machine-learning algorithms are also recognized tools in the general field of stock market prediction. In the literature of EWS, artificial neural networks \citep{kim2004korean, oh2006early, kim2004usefulness, yu2010multiscale, sevim2014developing, celik}, fuzzy inference \citep{Lin2008fuzzy, Nan2012fuzzy, giovanis2012study, fang2012adaptive} and support vector machines (SVM) \citep{hui2006research, hu2008financial, ahn2011usefulness} are proven accurate models for financial crisis prediction. Despite the promising accuracy demonstrated by those studies, few investigates the test-set early warning power of the model, that is the duration of forewarned period before the crisis onset.

To fill in the gaps discussed above, the objectives of this study are threefold. First, we attempt to develop a robust crisis classifier to precisely identify stock market turbulence on daily basis. The crisis classifier consists of two key components, namely the switching ARCH (SWARCH) model \citep{hamil} and two-peak (or valley-of-two-peaks) method \citep{rosenfeld1983histogram}. Instead of focusing on the return horizon, the proposed classifier tackles the problem from the perspective of the volatility \citep{rodtue, kim2013modeling, Fink2016, regimecopula2018, benmim2019financial}. The switching ARCH (SWARCH) model is adopted to label crisis/non-crisis episodes with high/low volatility regimes that imply market turbulence/tranquility \citep{hamil, hamlin, susram, susedw}. The model's effectiveness in depicting Chinese stock crises is explicitly examined in the authors' previous study on the contagion effect among housing, stock, interest rate and currency markets in China and the U.S. \citep{authors}. On the other hand, the two-peak method is an automatic thresholding approach \citep{jain1995machine} which selects classification thresholds automatically based on predetermined principles in order to obtain more robust segmentation. To classify stock turbulence, the two-peak method is performed on the histogram of SWARCH filtering possibilities to determine the optimal crisis cut-off. Second, a dynamic early warning system is developed integrating the crisis classifier and long short-term memory (LSTM) neural network \citep{jordan} to alert crisis onsets. As for the predictive model, LSTM is proven to be a state-of-art mechanism in the general field of financial forecasting \citep{chen2015lstm, fischer2018deep, wu2018adaboost, cao2019financial}, including volatility forecasting \citep{yu2018forecasting, kim2018forecasting, liu2019novel}. To the best of the authors' knowledge, this study is the first that incorporates LSTM in an EWS. Last, a comprehensive evaluation of the EWS is conducted by first examining the crisis classifier and predictor separately. To be specific, we empirically study the precision and robustness of the crisis classifier in comparison to the most widely used approach which defines stock crises according to an indicator functions of CMAX. The LSTM crisis predictor is then evaluated upon two baseline models, i.e. the back-propagation neural network (BPNN) and support vector regression (SVR), regarding to the performance metrics including the rand accuarcy, binary cross-entropy loss, receiver operating curve (ROC), area under curve (AUC) and the SAR score. To evaluate the effectiveness and stability of the EWS as a whole, the proposed algorithm is performed in not only the test set but also cross-validation and back-testing. According to the evaluation, the integrated EWS achieves the state-of-art performance and warns stock turbulence in the test set with $96.6\%$ accuracy and on average $2.4$ days ahead of crisis onsets. 

The remaining part of this paper is organized as follows. Section \ref{sec:data} describes the data included. Section \ref{structure} explicitly introduces the structure of the EWS and the algorithm related to the dynamic prediction of stock turbulence. Section \ref{sec:result} evaluates the model according to its performance, and Section \ref{conclusion} summarizes the conclusion. 

\section{Data}\label{sec:data}
\begin{table}[h!]
	\caption{Data description.}
	\setlength{\tabcolsep}{5pt}
	\centering
	\small
	\begin{tabular}{p{100pt}crr}
			\hline
			Data& 
			Frequency& 
			Reflection&Source\\
			\hline
			&&&\\
			Close price, log returns and realized volatilities of the SSEC index& 
			\vspace*{0.1mm}			
			Daily& 
			\vspace*{0.1mm}			
			Endogenous factors&
			\vspace*{0.1mm}			
		   WIND database\\	
			\hline		
			&&&\\
			S\&P500 Stock Price Index& 
			\vspace*{0.1mm}			
			Daily&
			\vspace*{0.1mm}			 
			US stock market&Yahoo finance\\
			\hline
			&&&\\
			USD/CNY exchange rate &Daily&Currency&US Federal Reserve Board\\
			\hline
			&&&\\
			Gold Price& 
			\vspace*{0.1mm}
			\multirow{2}{*}{Daily}& 
			\vspace*{0.1mm}
			\multirow{2}{*}{Global economy}&World Gold Council\\
			Oil Price	&
			\vspace*{0.1mm}
			&
			\vspace*{0.1mm}
			&International Monetary Fund\\
			\hline	
			&&&\\
			Interest rate for China(IMF published), M1, M2, CPI& 
			\vspace*{0.1mm}
			\multirow{4}{*}{Monthly}& 
			\vspace*{0.1mm}
			\multirow{4}{*}{Domestic economy} & \multirow{4}{*}{WIND database}\\
			\hline
	\end{tabular}
	\label{tab:data}
\end{table}

In this study, the Shanghai Stock Exchange Composite (SSEC) index is hired to reflect the Chinese stock market oscillation. Explanatory variables that are incorporated to predict stock crises are described in Table \ref{tab:data} in terms of frequency, purpose and source. Specifically, endogenous factors include the close price, log return and realized volatility\footnote{The realized volatility at time $t$ is defined as $\sigma_{rv}=\sqrt{\frac{1}{N_{t}}\sum_{t=1}^{N_{t}}(p_{t}-\bar{p_{t}})}$, where the $N_{t}$ is the count of days after time $t$, $p_{t}$ is the log return at $t$ and $\bar{p_{t}}$ is the average of log return til $t$.} of the SSEC index. The rest of the variables are exogenous factors of four genres reflecting the U.S. stock market, currency level, global and domestic economies, respectively. The samples span from Dec 27, 1998 to Oct 7, 2018 and are split into $70\%$ training and $30\%$ test sets. Table \ref{tab:stats_summary} shows the full sample statistics of the explanatory variables. 

\begin{table}[h!]
	\caption{Statistics of explanatory variables. St.Dev. is the standard deviation. $***$ and $**$ denote the (null normal) hypothesis test at the $1\%$ and $5\%$ significance level. $\dagger$ denotes the unit of M1 and M2 is $10^{13}$ Chinese yuan.}
	\setlength{\tabcolsep}{5pt}
	\centering
	\small
	\begin{tabular}{p{110pt}rrrrr}
		\hline
		& 
		Mean& 
		St.Dev. & Skewness & Kurtosis & Jarque-Bera \\
		\hline
		&&&&&\\
		SSEC Close Price&2766.65&560.77&0.68&1.01&$291.46^{***}$\\
		SSEC log return&0.02&1.49&-0.78&4.86&$2643.2^{***}$\\
		SSEC realized volatility&1.7&0.31&1.86&4.05&$3069.1^{***}$\\
		S\&P500 Index&1682.81&529.84&0.19 & -1.04&$124.03^{***}$\\
		USD/CNY exchange rate&6.49&0.27&0.06&-1.46&$217.38^{***}$\\
		Gold Price&1296.08&231.33&0.24&-0.14&$26.129^{***}$\\	
		Oil Price&73.25&22.88&-0.12&-1.41&$208.00^{***}$\\
		Interest rate for China&3.06&0.22 &0.73&2.22&$717.48^{**}$\\
		M1&$3.38^{\dagger}$&$1.08^{\dagger}$&0.47&-0.79&$151.87^{***}$\\
		M2&$1.10^{\dagger}$&$3.91^{\dagger}$&0.12&-1.21&$154.91^{***}$\\
		CPI&95.83&6.78&-0.4  &-1.04&$174.59^{***}$\\
		\hline
	\end{tabular}
	\label{tab:stats_summary}
\end{table}

\section{An integrated early warning model}\label{structure}
\subsection{Crisis identification with SWARCH and two-peak method}\label{subsec:EWS}
\subsubsection{High/low volatility regimes in the stock oscillation}\label{subsubsec:SWARCH}
Stock crashes are inevitable results of volatility jumps. To explain this phenomenon, we propose to investigate the high/low volatility regime of the stock return based on the SWARCH model \citep{hamil}. The target is to provide a reliable solution to crisis warning from the perspective of risk.

Following \cite{hamil}, the log return of stock price with high/low volatility regimes could be formulated as a AR(1)-SWARCH(2,1) process given by:
\begin{align}
y_{t} &= u + \theta_{1}y_{t-1}+\epsilon_{t},\quad \epsilon_{t}|\mathcal{I}_{t-1}\sim N(0, h_{t});\label{swarch1}\\
\frac{h_{t}^{2}}{\gamma_{s_{t}}} &= \alpha_{0}+\alpha_{1}\frac{\epsilon_{t-1}^{2}}{\gamma_{s_{t-1}}}, s_{t}=\{1,2\}.\label{swarch2}
\end{align}
Eq.(\ref{swarch1}) describes an AR(1) process with a normal error term $\epsilon_t$ of variance $h_{t}$. The regime switching structure of the residual variance $h_{t}$ is given by Eq.(\ref{swarch2}) where the $\alpha's$ are non-negative, the $\gamma's$ are scaling parameters that capture the change in each regime, $s_{t}$ is the state variable that $s_{t}=1$ indicates the low volatility state, and $s_{t}=2$ indicates the high volatility state.

The probability law which results in the stock market switching between the high/low volatility regimes is assumed to be the constant transition probabilities of a two-state Markov chain,
\begin{align}
p_{ij} = Prob(s_{t}=j|s_{t-1}=i), i,j = \{1,2\}. 
\end{align}

The classification of high/low volatility regimes can be implemented on the basis of the filtering probability, which is a byproduct of the maximum likelihood estimation. The filtering probability based on historical observations till time $t$, $Y_{t}$, written as
\begin{align}\label{eq:filt}
P(s_{t}=i|Y_{t};\boldsymbol{\theta}_{t})
\end{align}
where $\boldsymbol{\theta}_{t}$ is the vector of model parameters to be estimated. Given that $s_{t}=2$ is the state of high volatility, $P(s_{t}=2|Y_{t};\boldsymbol{\theta}_{t})$ can be interpreted as the conditional probability of crises based on the current information of time $t$. We thus define stock turbulence as the following binary function.

\begin{align}\label{eq1}
\text{Crisis}_{t} = 
\begin{cases}
1, & P(s_{t}=2|Y_{t};\boldsymbol{\hat{\theta}}_{t}) \geq c \\
0, & \text{otherwise.}
\end{cases} 
\end{align}	
where $\boldsymbol{\hat{\theta}}_{t}$ is the estimated parameter vector and $c$ is the crisis threshold/cutoff point.

In this way, stock crisis classification is structured through the mechanism that filtering probabilities of the system being in the high volatility regimes tend to increase as the stock price becomes more volatile, and there exists a threshold $c$ which identifies crises once it is exceeded. By Eq.\ref{eq1}, $c$ indicates the lowest-level likelihood of the high-volatility state that could be considered as crises. Hence the determination of $c$ plays a key role in the EWS. 

\subsubsection{Crisis thresholding: two-peak method}\label{subsubsec:threshold}
To balance the trade-off between sensitivity and false alarms \citep{babecky2014banking}, this study adopts the two-peak method to automatically determine crisis thresholds. The two-peak method is developed with the general purpose of finding the optimal threshold in the context of binary classification, and is proven experimentally credible in solving image processing-related classification problems \footnote{\cite{Prewitt1966} first introduce the two-peak method in the cell image analysis of distinguishing the gray-level difference between the background and the object. The performance of the method is further verified in \cite{rosenfeld1983histogram} by analyzing the histogram's concavity structure.}. According to the two-peak method, the optimal threshold of a binary system is the minimum value between the two peaks of the frequency density histogram \citep{Weszka1978A}. There are several alternative thresholding mechanisms that are built on the histogram, such as the Otsu's method \citep{Ohtsu2007A} that solves the multi-threshold problem by considering the pixel variance. In this study, we use two-peak as it is the most straightforward of all, and the foundation of other approaches thereafter. 

Given that our crisis classifier has two state classes, i.e. crisis (1) and non-crisis (0), the two-peak method is applied to determine the crisis cutoff based on the SWARCH filtering probabilities of the high-volatility state $P(s_{t}=2|Y_{t};\boldsymbol{\hat{\theta}}_{t})$. Specifically, we first sketch the histogram of high-volatility filtering probabilities from time $0$ to $t$. The valley bottom between the two frequency peaks is then selected as the optimal cutoff point at $t$. To further enhance the robustness of our system, the two-peak method is performed on a recursive basis to obtain dynamic thresholds as the prediction moves forward (See Algorithm \ref{algorithm} in the next section).

\subsection{Crisis warning with long-short term memory neural network}\label{subsec:LSTM}
The long-short term memory (LSTM) network \citep{jordan} belongs to the family of recurrent neural networks (RNNs) \citep{hochreit} and is designed to learn both long- and short-term dependencies for sequential forecasting. As a deep learning model, LSTM networks nowadays are widely used in the financial sector in a variety of areas from stock prediction to risk management. 

\begin{figure}[h]
	\centering
    \includegraphics[width=0.8\textwidth]{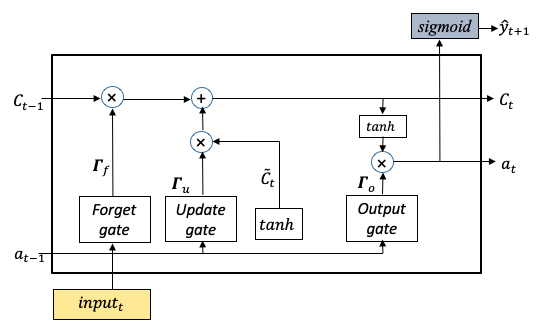}
    \caption{The LSTM cell inner structure at time $t$.}\label{fig:lstm_cell}
\end{figure}

As an extension of classic RNNs, LSTM keeps its merit to allow the processing of sequential data with arbitrary lengths via the hidden state vector, at the same time enhances the learning power of long-distance dependency by introducing the so-called memory cell. As it is displayed in Figure \ref{fig:lstm_cell}, the inputs of a LSTM cell at time $t$, namely $a_{t-1}$ and $C_{t-1}$, are memories that contain historical information passed through from the former cell in the form of activation and peephole functions. $\mathbf{\Gamma}_{f}$, $\mathbf{\Gamma}_{u}$, $\mathbf{\Gamma}_{o}$ are sigmoid functions of the forget gate, the update gate and the output gate that determine the information to be discarded, added and reproduced, respectively. $\tilde{C}_{t}$ is the new candidate output created by the $tanh$ layer, which is limited in the range $[-1,1]$. Finally, three outputs, $\hat{y}_{t+1}$, $a_{t}$ and $C_{t}$, are produced for the current cell at time $t$, where $a_{t}$ and $C_{t}$ are recurrently employed as the inputs of the next memory block\footnote{The initial values of $C_{0}$ and $a_{0}$ are both zero.}. Note that the last sigmoid function in the upper right corner is only included in the last cell of the LSTM network, and is used to produce the network output $\hat{y}_{t+1}$ in $[0,1]$.

\begin{figure}[h]
    \centering
	\includegraphics[width=\textwidth]{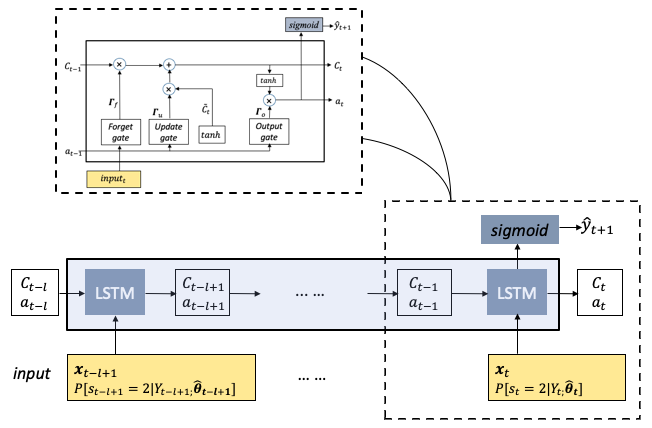}
	\caption{LSTM with window size $l$. The LSTM cell structure in Fig. \ref{fig:lstm_cell} is the last cell of the window.}
	\label{fig:lstm_struct}
\end{figure}

For each cell of LSTM, the formulae of the three gates, $\mathbf{\Gamma}_f, \mathbf{\Gamma}_{u}, \mathbf{\Gamma}_{o}$ and the new candidate state $\tilde{C}_{t}$ can be written as:
\begin{align*}
	\mathbf{\Gamma}_{f} &= \sigma(x_{t}U^{f} + a_{t-1}W^{f});\\
	\mathbf{\Gamma}_{u}&=\sigma(x_{t}U^{u} + a_{t-1}W^{u});\\
	\mathbf{\Gamma}_{o}&=\sigma(x_{t}U^{o} + a_{t-1}W^{o});\\
	\tilde{C}_{t}& = tanh(x_{t}U^{g}+a_{t-1}W^{g})
\end{align*}
where $\sigma$ is the sigmoid function, $x_{t}$ is the input vector, $a_{t}$ is the activation, $U$ is the weighted matrix connecting inputs to the current layer, $W$ is the recurrent connection between the previous and current layers. Therefore, $\mathbf{\Gamma}_{f,u,o}$ implies the level of information that each gate processes after balancing between the previous activation and the current input. The candidate state $\tilde{C}_{t}$ is computed based on the current input and the previous hidden state, and later added to the next cell state $C_{t}$ on the basis of $C_{t-1}$. 

This study applies LSTM as the predictive model and infers stock market turmoils on daily basis using historical information of a fixed window size $l$. As Figure \ref{fig:lstm_struct} shows, each prediction is made from a network of $l$ LSTM memory blocks that sequentially process the input of both the explanatory variables $\{\boldsymbol{x}_{t-l+1},...,\boldsymbol{x}_{t}\}$ and the SWARCH filtering probability $\{P[s_{t-l+1}=2|Y_{t-l+1};\boldsymbol{\hat{\theta}}_{t-l+1}]$,...,$P[s_{t}=2|Y_{t};\boldsymbol{\hat{\theta}}_{t}]\}$ from time $t-l+1$ to $t$, for $t \geq l$. The output $\hat{y}_{t+1}$ is produced by a sigmoid function indicating the probability of high-volatility at $t+1$. Early warning signals are thus released for time $t+1$ once the value of $\hat{y}_{t+1}$ exceeds the two-peak threshold at $t$ (See Section \ref{subsubsec:threshold}). The LSTM network consists of $13$ input layers (the number of the input variables), $32$ LSTM layers and the output layer, which brings $5921$ parameters to be trained. The batch size and epoch number are $20$ and $100$, respectively. Given the sample size of $T$ days, $T-l$ predictions will be made from $t=l+1$ onward.

\begin{figure}[!h]
	\centering{\includegraphics[width=0.9\textwidth]{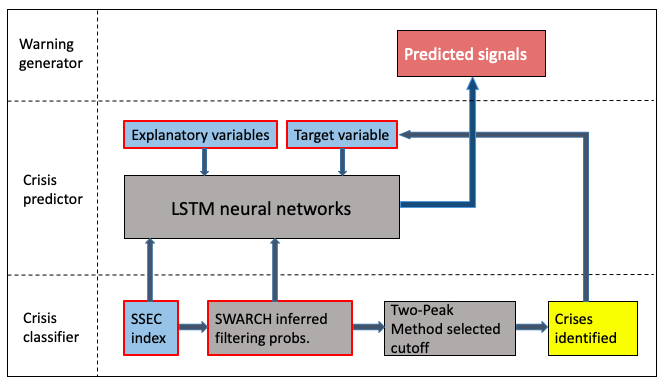}}
	\caption{The structure of EWS with the crisis classifier, crisis predictor and the warning generator.}
	\label{fig:ews}
\end{figure}

Figure \ref{fig:ews} structures the integrated EWS regarding to its three key components, i.e. the crisis classifier, crisis predictor and warning generator. Specifically, the crisis classifier identifies stock market turmoils according to Eq. \ref{eq1} based on the SWARCH filtering probability and the crisis cutoff determined by the two-peak method. The output of the crisis classifier then becomes the target variable and is fed into the LSTM crisis predictor together with other explanatory variables. Finally, early warning signals are generated as the predicted output exceeds the crisis cutoff. To make robust daily predictions, the system is performed on a dynamically-recursive basis. The procedure is described by Algorithm \ref{algorithm} on the sample of size $T$.

\IncMargin{1em} 
\begin{algorithm}[!h]	
	\SetAlgoNoLine 
	\SetKwInOut{Input}{\textbf{Initial inputs}}\SetKwInOut{Output}{\textbf{Final output}} 
	
	\Input{
		\\
		The SSEC index price $P_{t}$\;\\
		The explanatory variables $\boldsymbol{x}_{t}$ excluding $P_{t}$\;\\}
	\Output{
		\\
		The predicted signals $\hat{y}_{t+1}$\;\\}
	\BlankLine
	
	calculate log returns of SSEC index price, $\{logR_{t}, t=1,...,T\}$\; 
	set up the window size $l$\;
{\For {$t$ from $l$ to $T$}{
    \Repeat
{\text{t=T}}
{\For{$i$ from $1$ to $t+1$}{input $logR_{i}$ into SWARCH\; 
		output filtering probability $P[s_{i}=2|Y_{i};\boldsymbol{\hat{\theta}}_{i}]$\;}
	two-peak method selects the optimal cutoff $c_{t}$\;
	\For{i from $1$ to $t+1$}{
	{\If{$P[s_{i}=2|Y_{i};\boldsymbol{\hat{\theta}}_{i}]\geq c_{t}$}{
   Crisis$_{i}=1$\;}}}	
    \For {$j$ from $1$ to $l$}{
		input explanatory variable vector $\boldsymbol{x}_{t-l+j}$, filtering probability $P[s_{t-l+j}=2|Y_{t-l+j};\boldsymbol{\hat{\theta}}_{t-l+j}]$ and identified crisis signals Crisis$_{t-l+j}$ into LSTM\;

        }
    	output the prediction $\hat{y}_{t+1}$\;
    }
}
}
      \caption{Daily warning for Chinese stock turbulence}	
      \label{algorithm}
\end{algorithm}
\DecMargin{1em}

\section{System evaluation}\label{sec:result}
In this section, a comprehensive evaluation is conducted by studying first the crisis classifier and predictor (see Figure \ref{fig:lstm_struct}) separately, then the early warning system as a whole. In the view of the crisis classifier that jointly uses the SWARCH and two-peak method, we intend to understand its precision and robustness with empirical evidences. Next, the LSTM predictor is evaluated with two baselines, i.e. the back-propagation neural network (BPNN) and support vector regression (SVR), according to the performance metrics consisting of the rand accuracy \citep{rand}, binary cross-entropy loss \citep{Shannon1948A}, receiver operating curve (ROC), area under curve (AUC)\citep{roc} and the SAR score \citep{sar2004}. Last, the early warning power of the entire system is investigated according to its test-set performance, cross-validation as well as back-testing.

\subsection{Evaluating the crisis classifier}
The credibility of an EWS is rooted in a precise and robust crisis classifier. According to Figure \ref{fig:lstm_struct} and Algorithm \ref{algorithm}, stock crisis cutoffs are computed dynamically for each prediction taking into account the current market condition as well as past information. To validate the reliability of the proposed classification mechanism, we analyze the crisis identification results in terms of its precision and robustness. 

\begin{figure}[h]
	\centering
    \includegraphics[width=0.95\linewidth]{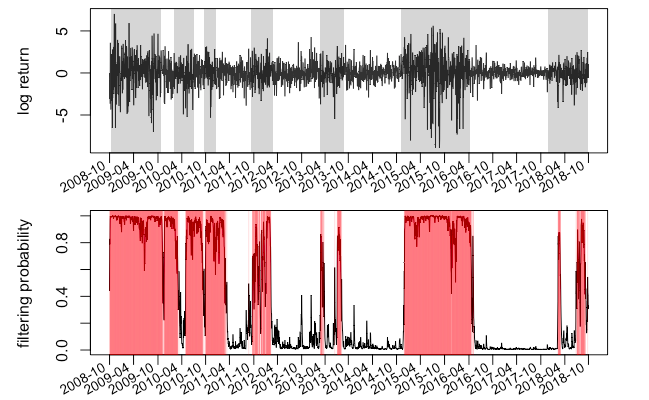}
	\caption{Log return of the SSEC stock index (upper panel) and the corresponding high-volatility filtering probability (lower panel). Turmoil periods determined by Algorithm 1 are highlighted in grey and red.}
	\label{fig:filtering_prob}
\end{figure}

As crisis classification is a subjective topic heavily depending on the individual understanding of crisis, limited analysis could be done on quantitatively evaluating the accuracy due to the lack of true crisis labels. Given the target of the proposed EWS is to predict stock market turbulence, we investigate the precision of the crisis classifier with emphasis on the empirical evidence related to volatility regimes. Figure \ref{fig:filtering_prob} and Table \ref{crises} summarizes the turmoils classified in the Chinese stock market by performing Algorithm \ref{algorithm} on the full sample. In Figure \ref{fig:filtering_prob}, crisis periods are highlighted in both the log return (grey in the upper panel) and filtering probability plots (red in the lower panel). As Figure \ref{fig:filtering_prob} suggests, the proposed hybrid algorithm captures all the recorded stock crises that are also reflected by volatile log return and filtering probability jumps. Table \ref{crises} lists the starting and ending days of the detected turmoils with their associated critical events. The hybrid classifier identifies crises with promising results explaining not only major global turmoils including the 2008 global financial crisis and 2010 European debt crisis, but also local stock turbulence resulted from the industrial reformation in 2013, the high-leveraging bubble collapse in 2015 and the economic slowdown since 2018. 

\begin{table}[h!]
    \caption{Turmoil periods that are identified by Algorithm 1 in the full sample and associated critical events.}
	\setlength{\tabcolsep}{5pt}
	\centering
	\small
\begin{tabular}{lcc}
    \hline
    Event & Identified Crisis Period\\
    \hline
    \multirow{2}{*}{2008 Global financial crisis}& 2008/10/04 - 2009/11/06 \\
   & 2009/11/16 - 2010/03/28  \\
    \hline
    \multirow{3}{*}{2010 European debt crisis}
   & 2010/05/06 - 2010/09/16 \\
   & 2010/10/08 - 2011/03/17  \\
   & 2011/09/22 - 2012/02/17 \\
   \hline
   \multirow{1}{*}{2013 Industrial reformation}&   2013/03/04 - 2013/08/12 \\
    \hline
    \multirow{2}{*}{2015 Chinese stock crash}& 2014/12/02 - 2016/04/27 \\
    & 2016/05/09 - 2016/05/11 \\
    \hline
    \multirow{4}{*}{2018 Domestic economy slowdown}& 2018/02/09 - 2018/03/06 \\
   & 2018/07/02 - 2018/08/03  \\
   & 2018/08/06 - 2018/08/31 \\
   & 2018/09/04 - 2018/09/26 \\
    \hline
\end{tabular}\label{crises}
\end{table}

\begin{figure}[h]
	\centering
	\includegraphics[width=0.9\linewidth]{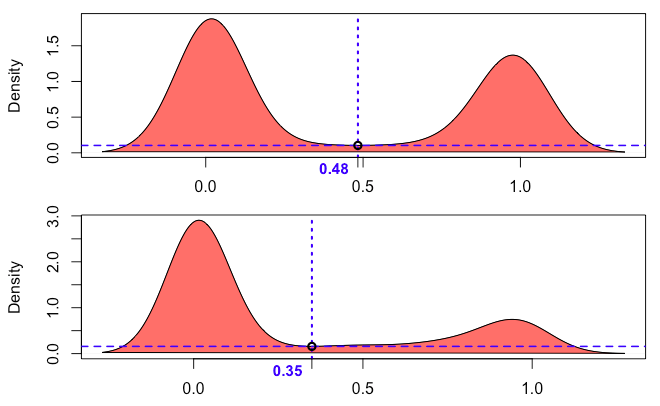}
	\caption{Cutoffs selected by the two-peak method in the full sample (upper panel) and test set (lower panel).}
	\label{fig:cutoff}
\end{figure}

The robustness of a model broadly refers to its error-resisting strength and resilience in producing results as data changes. Therefore, robust crisis classifications are subject to a dynamical thresholding mechanism to handle turbulence with limited influence from sample variations. Table \ref{tab:cutoffs} summarizes the statistics of crisis cutoffs that are determined in the full sample and test set by Algorithm \ref{algorithm}. The number of cutoffs in a sample is given by the difference between the number of observations $T$ and the window size $l$. With windows of size $5$ (days), this study computes $2430$ and $725$ cutoffs in the full sample and test set of lengths $2434$ and $729$ (days), respectively. As Table \ref{tab:cutoffs} displays, the cutoff distributions of the full sample and test set are both right skewed given the greater means ($0.515$, $0.429$) than the medians ($0.489$, $0.396$) and modes ($0.483$, $0.355$). In other words, the positive skewness indicates that cutoffs are more likely to take values below the mean and around the median/mode. Moreover, test-set cutoffs exhibit lower values with mean, median and mode approximating to $0.4$, whereas those in the full sample are closer to $0.5$. To explain this difference in the crisis cutoff distributions, Figure \ref{fig:cutoff} shows the smoothed histograms of SWARCH filtering probabilities in the full (upper panel) and test (lower panel) sets. The optimal cutoffs determined at the end of Algorithm \ref{algorithm} for the last day observation are circled in blue. Although the test set exhibits a greater proportion of tranquil days with a significantly higher right peak, the two-peak method detects the true valley at $0.35$ to threshold the crisis.

\begin{table}[h]
   \setlength{\tabcolsep}{10pt}
    \centering
    \caption{Statistics of crisis cutoffs in the full sample and test set.}\label{tab:cutoffs}
    \begin{tabular}{lcccccc}
    \hline
       & Count & Mean & St.Dev & Median & Mode & Range\\ \cline{2-7}
       &&&&&&\\
    $\text{Cutoff}_{\text{full-sample}}$ & 2430 & 0.515 & 0.128 & 0.489 & 0.483 & 1.00 \\
    $\text{Cutoff}_{\text{test-set}}$ & 725 & 0.429 & 0.121 & 0.396 & 0.355 & 0.996 \\
    \hline
    \end{tabular}
\end{table}

 Further with the argument that a robust classification model ought to produce stable classification results regardless of the sampled information, Table \ref{cmax} compares stock crises identified by Algorithm \ref{algorithm} with those defined on the CMAX indicator\footnote{The CMAX index is the most widely used crisis indicator in the literature concerning stock market early warning \citep{coudert2008does, li2015toward, fu2019predicting}. It defines stock crashes with an indicator function $1_{CMAX_t<\mu_t-\lambda\sigma_t} CMAX_t:1$, where $\mu_t$ and $\sigma_t$ are the mean and standard deviation of $CMAX_t$, and $\lambda$ is a market-dependent constant \citep{klr}. In this study, we consider four cases when $\lambda=1, 1.5, 2, 2.5$ as they give reasonable results for Chinese stock market crises.}. Daily classifications are computed in both the full-sample and test set for each model. To examine the level of consistency between crises identified on different samples, Table \ref{cmax} lists the number (Row 3) and percentage (Row 5) of days that the full-sample crises differ from the test-set crises during the period from 2015/10/13 to 2018/09/28 (729 days in total)\footnote{This is the period when full sample and test set intersect.}. With $16$ days of deviation in a period of almost three years and a percentage of $2.19\%$\footnote{We believe that the percentage deviation of $2.19\%$ could be further reduced with a larger sample of test set and cross validation. Relevant analyses on this aspect will be conducted in the future study.}, the integrated EWS produces the most robust crisis classification result in comparison to the CMAX indicator on a range of parameters $\lambda=1, 1.5, 2, 2.5$.

\begin{table}[!h]
 \caption{Difference between crises identified on the full sample and test set during 2015/10/13 - 2018/09/28.}
	\setlength{\tabcolsep}{2pt}
	\centering
	\small
\begin{tabular}{lccccc}
\hline
 & Integrated EWS & CMAX$_{\lambda=1}$ & CMAX$_{\lambda=1.5}$ & CMAX$_{\lambda=2}$ & CMAX$_{\lambda=2.5}$\\
 \cline{2-6}
 &&&&&\\
No. of crises with full sample & 191 & 203 & 148 & 3 & 0\\
No. of crises with test set  & 207 & 154 & 112& 115 & 67 \\
No. of non-matching days & $\boldsymbol{16}$ & 49 & 36 & 112 & 67 \\
Total no. of days & 729 & 729 & 729 & 729 & 729 \\
\% of non-matching days & $\boldsymbol{2.19}$ & 6.27 & 4.94 & 15.4 & 9.19 \\
\hline
\end{tabular}\label{cmax}
\end{table}

\subsection{Evaluating the crisis predictor}\label{subsec:model_compare}
We now evaluate the crisis predictor based on LSTM in comparison to two baselines of BPNN and SVR. The associated performance metrics is discussed in Section \ref{subsec:perf_measure}. And Section \ref{subsec:perf_outs} presents the results. 

\subsubsection{Evaluation metrics}\label{subsec:perf_measure}

The evaluation metrics of the predictor include three classes of performance measures that are designed for classification models, i.e. (I) the rand accuracy (\citeauthor{rand}, \citeyear{rand}) and binary cross-entropy loss (\citeauthor{Shannon1948A}, \citeyear{Shannon1948A}), (II) the receiver operating curve (ROC) and area under curve (AUC) \citep{roc}, and (III) the SAR score \citep{sar2004}. Prior to the performance evaluation, Table \ref{tab:truefalse} lists the confusion matrix that is used by the rand accuracy, ROC and SAR score. 

\begin{table}[h]
  \centering
  \caption{Confusion matrix for daily stock early warning.}
  	\setlength{\tabcolsep}{8pt}
    \begin{tabular}{|c|c|c|}
    \hline
    Actual/Predicted & 1: Crisis & 0: Non-crisis \\
    \hline
    1: Crisis     & True positive (TP) & False negative (FN) \\
    \hline
    0: Non-crisis     & False positive (FP) & True negative (TN) \\
    \hline
    \end{tabular}%
  \label{tab:truefalse}%
\end{table}%

In general, true positive/negative corresponds to the true prediction of turmoil/tranquility, whereas false positive/negative corresponds to the false prediction. Moreover, the true positive rate (TPR) and false positive rate (FPR) are defined as the percentage of truly predicted crisis signals over the total number of actual crises, and the percentage of falsely predicted crisis signals over the total number of actual tranquility, respectively.

\begin{align}
\text{TPR} = \frac{\text{TP}}{\text{TP + FN}}, \quad 
\text{FPR} = \frac{\text{FP}}{\text{FP + TN}}. 
\end{align}

Evaluation Metric I: The rand accuracy is defined as the proportion of true results over the total number of cases examined:

\begin{align}\label{eq4}
\text{Accuracy} = \frac{\text{TP + TN}}{\text{TP + TN + FP + FN}}.
\end{align}

The binary cross-entropy loss measures the performance of classification models in terms of the level that the predicted probability of getting $1$ deviates from the true label $0$ or $1$, and is expressed as:

\begin{align}\label{eq5}
\text{Loss} = -\frac{\sum_{i=1}^{n-l+1}(y_{i} log(\hat{y_{i}})+(1-y_{i})log(1-\hat{y_{i}}))}{n-l+1},
\end{align}

where $y_{i}$ and $\hat{y}_{i}$ denote the true and predicted values, and $n$ is the sample size. As we set the label of crises to be $True$ ($=1$), an EWS model that warns all the crises regardless the number of $False$ alarms it creates, has zero loss indicating none of the crisis is lost. According to Eq. (\ref{eq4}) and (\ref{eq5}), a greater level of predictive power comes along with higher rand accuracy and lower binary cross-entropy loss.

Evaluation Metric II: As one of the most classic performance measures, ROC plots the FPR (x-axis) against the TPR (y-axis) for each classifier. As a higher true positive rate is always more preferable given the level of the false positive rate, models with the ROC curve bending closer towards the upper-left corner are more preferable. To offer a quantitative representation of the graphic information carried by ROC, AUC computes the total area under the ROC curve and suggests the better model with the greater AUC value.

Evaluation Metric III: Different from the widely-used F1-score, the SAR score \citep{sar2004} is developed as a more holistic performance measure due to the uncertainty of the correct evaluation metric. By taking into account three distinctive measures including the accuracy, AUC and root mean-squared error (RMSE), models with higher SARs are regarded as better-performing as they produce overall high accuracy/AUC and low RMSE. 

\begin{align}\label{eq6}
  \text{SAR} = \frac{1}{3}(\text{Accuracy} + \text{AUC} + (1- \text{RMSE})). 
\end{align}

\subsubsection{Test-set performance}\label{subsec:perf_outs}
 To evaluate the predictive power of LSTM, BPNN and SVR, Table \ref{tab:acc_loss} preliminarily lists the test-set rand accuracy and binary cross-entropy loss of the three models following Algorithm \ref{algorithm}\footnote{To obtain the baseline results, Algorithm \ref{algorithm} is implemented by replacing the LSTM in line 16 by BPNN and SVR.}. Three window sizes $l= {22, 10, 5}$ are considered. As Table \ref{tab:acc_loss} suggests, LSTM with window size $l=5$ produces the optimal crisis prediction that yields the highest accuracy $0.952$ and lowest loss $0.27$ among all cases examined. Among the three predictive models, LSTM consistently demonstrates the strongest forecasting power of stock crises given different window sizes. Moreover, it is observed that with the last five days of information, all the three models achieve the best result (except the accuracy of SVR) in comparison to the predictions made with $22$ and $10$ days information. Therefore, the remaining of the evaluation is conducted with window size $5$. 
 
\begin{table}[h!]
	\caption{Test-set rand accuracy and binary cross-entropy loss based on LSTM, BPNN and SVR with varying the window sizes}\label{tab:acc_loss}
	\setlength{\tabcolsep}{15pt}
	\centering
	\small
	\begin{tabular}{lccc}
			\hline
			&&&\\
			 &
			 LSTM &
			 BPNN & 
			 SVR
			\\ \cline{2-4}
			Window size $l=22$ &&&\\ \cline{1-1}
			&&&\\
			Accuracy& 	
			0.930& 	
			0.882 & 0.927\\			
			Binary cross-entropy loss& 		
			0.380&	 
			 0.439& 0.407\\
			\hline	
			&&&\\
			Window size $l=10$ &&&\\ \cline{1-1}
			&&&\\
			Accuracy& 	
			0.941& 	
			0.865 & 0.920\\			
			Binary cross-entropy loss& 		
			0.326&	 
			0.305& 0.405\\
			\hline
			&&&\\
		    Window size $l=5$ &&&\\ \cline{1-1}
			&&&\\
			Accuracy& 	
			\textbf{0.952}& 	
			0.899 & 0.912\\			
			Binary cross-entropy loss& 		
			\textbf{0.270}&	 
			0.369& 0.423\\
			\hline
	\end{tabular}
\end{table}
 
\begin{figure}[H]
	\centering 
	\subfloat[ROC]{\includegraphics[width=0.5\linewidth]{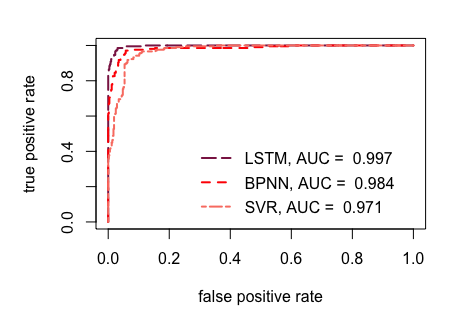}}
	\subfloat[LSTM]{\includegraphics[width=0.5\linewidth]{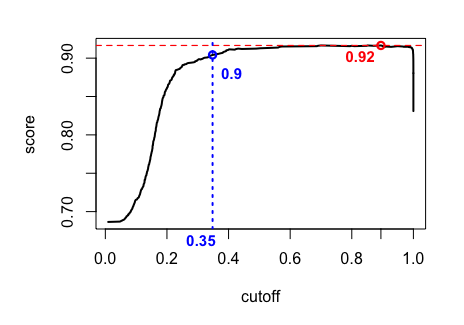}}\\
	\vspace*{-3mm}
	\subfloat[BPNN]{\includegraphics[width=0.5\linewidth]{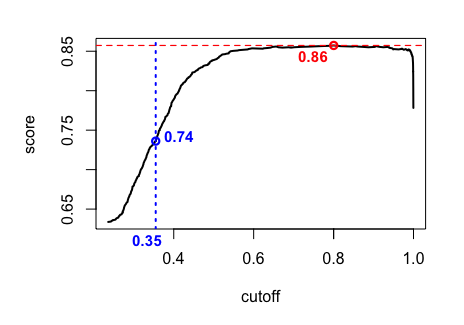}}
	\subfloat[SVR]{\includegraphics[width=0.5\linewidth]{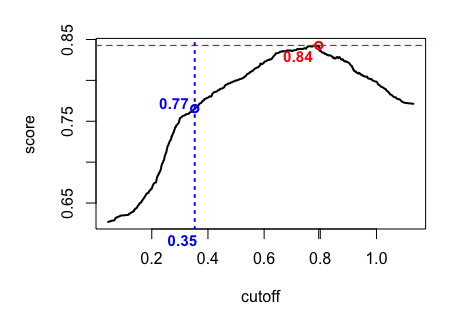}}
	\caption{Test-set ROC (Panel a) and SAR (Panel b-d) curves of LSTM, BPNN and SVR.}
	\label{fig:roc}
\end{figure}

 Figure \ref{fig:roc} further shows the test-set ROC and SAR curves. In particular,
 Panel (a) shows the ROC curves and AUC values generated from the test-set predictions. As the ROC-oriented metric tells the model's ability in classifying the binary states, LSTM enhances BPNN and SVR with its outstanding capacity in distinguishing turbulence/tranquility with the optimal ROC curve and AUC value of $0.997$. 
 
 Panel (b-d) plot the SAR score against the crisis cutoff for the three predictive models. According to Algorithm \ref{algorithm}, the test-set score of each model is highlighted as the blue point in each panel corresponding to the last day cutoff obtained from the dynamic crisis classifier, whereas the red point is the highest score obtained by the predictive model regardless of the optimal cutoff.  From the perspective of model scores, LSTM remains its dominating state with the highest test-set score (blue) of $0.9$, whereas BPNN and SVR score $0.74$ and $0.77$, respectively. Moreover, LSTM appears to be the most insensitive model to cutoff variations as the scores remain relatively high in a prolonged range shaped as a flat peak in Panel (b). With a similar shape in Panel (c), BPNN produces a SAR curve with reduced scores and a smaller peak, where the test-set score $0.74$ exhibits a large deviation from the best score of $0.86$. Despite that SVR produces close scores as BPNN, the sharp peak in Panel (d) suggests the model's instability in predicting with varying cutoffs. 
 
\subsection{Crisis early warning}\label{subsec:model_comp}
In this section, we examine the integrated EWS in terms of its early warning power with respect to the forewarned period ahead of the actual crisis onsets. By keeping BPNN and SVR as baselines, test-set forecasting, cross-validation and back-testing are implemented. In this way, we hope to gain a comprehensive understanding on the system's crisis forecasting capacity, stability as well as effectiveness. 

\subsubsection{Test-set performance}
\begin{figure}[!h]
	\centering
	    \includegraphics[width=0.7\textwidth]{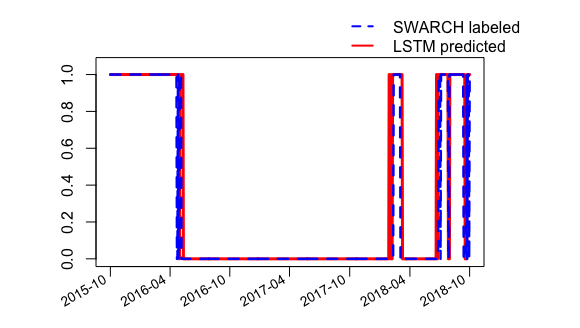}\\ 
        \includegraphics[width=0.7\textwidth]{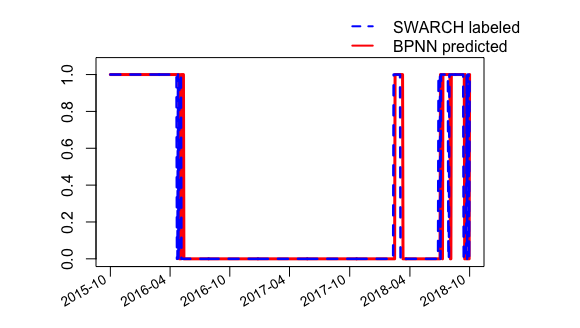}\\
    	\includegraphics[width=0.7\textwidth]{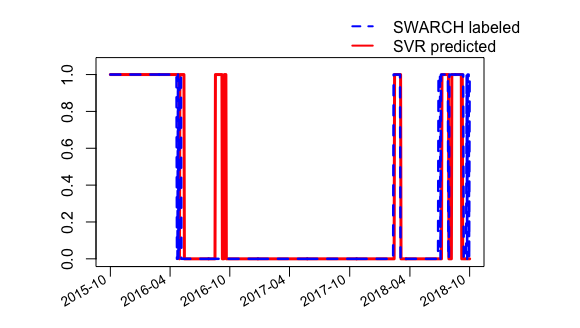}
	\caption{Test-set early warning signals }
	\label{fig:pred} 
\end{figure}

Figure \ref{fig:pred} shows the predicted signals by the integrated EWS against their true crisis labels ($1$ for crisis and $0$ otherwise) by the SWARCH model. As Figure \ref{fig:pred} displays, crisis onsets in the test set mainly occur in 2016 as a result of the lasting effect from the 2015 stock market crash, and in 2018 due to the financial instability in China. Overall, the proposed EWS with LSTM predictions depict the test-set set crises in a relatively precise manner with the first alarms (red line) before the actual onsets (blue dashed line). As the predictive model is replaced by BPNN, the EWS tends to delay in producing the first crisis signal despite of its ability in capturing ongoing crises. In contrast to LSTM and BPNN, SVR appears to suffer from both delayed warnings and false alarms in Figure \ref{fig:pred}. 

To support the preceding claims with evidence, Table \ref{tab:correct_pred} summarizes the numerical results related to the test-set forecasting. The test set consists of $729$ days with $207$ crisis days (Row 2, Table \ref{tab:correct_pred}) and $6$ crisis onsets (Row 6, Table \ref{tab:correct_pred}). With respect to Table \ref{tab:correct_pred}, EWS with LSTM demonstrates a promising capability of warning stock turbulence that is reflected by its dominating results in all aspects examined. In particular, LSTM-based EWS improves the baselines with $200$ days of correct predictions which yield a rate of $96.6\%$. On average, the model alerts stock turbulence $2.4$ days ahead of the actual crises and successfully warns $83.3\%$ of the onsets with $0\%$ false alarm. It is worth-mentioning that the missed onset occurs two days after its preceding crisis on July 25, 2018 and lasts for one day only. In line with the observations made from Figure \ref{fig:pred}, the major weakness of the BPNN-based EWS reveals due to its delay in generating crisis signals, which is suggested by a relatively high rate of correct daily predictions $94.6\%$ and a low rate of successfully predicted onsets $33.3\%$. Beside the delays, the high percentage of $30\%$ false alarms makes SVR the least reliable model for the early warning task in comparison to LSTM and BPNN. 

\begin{table}[!h]
	\caption{Summary of test-set forecasting. \% of correct predictions is the percentage of correctly predicted crisis signals, \% of correct predicted onsets is the percentage of correctly forewarned onsets. }	
	\setlength{\tabcolsep}{15pt}
	\centering
	\small
    \begin{tabular}{cccc}
            \hline
            Model &LSTM&BPNN&SVR\\ \hline
            &&&\\
            Total crises & 207 & 207 & 207\\
           	Correct predictions & \textbf{200}&196&184\\
            \% of correct predictions&\textbf{96.6}&94.6&88.9\\
            \hline
            &&&\\
            Total onsets&6&6&6\\
            Predicted onsets&\textbf{5}&2&2\\
            \% of correct predicted onsets&\textbf{83.3}&33.3&33.3\\
            \% of false onset alarms&0.0&0.0&30.0\\
            Avg. days-ahead onsets &\textbf{2.4}&1.5&2.0\\
      	  	\hline
	\end{tabular}\label{tab:correct_pred}	
\end{table}

\subsubsection{Cross validation}
To analyze the stability of the EWS, a $k$-fold cross validation is further conducted in the test set with varying values $k=3,5,8$\footnote{Given the selection of $k$ deals with the trade-off between bias and variance, the cross validation is conducted up to $8$ folds in order to ensure the size of the test set is large enough to offer statistically representative of the model's forecasting power.}. Rand accuracy and cross-entropy loss are used as the performance measures.

\begin{table}[h!]
	\caption{Average test-set rand accuracy and binary cross-entropy loss from the $k$-fold cross validation}	
	\setlength{\tabcolsep}{15pt}
	\centering
	\small
	\begin{tabular}{lccc}
			\hline
			&&&\\
			 &
			 LSTM &
			 BPNN & 
			 SVR
			\\ \cline{2-4}
			$k=3$&&&\\ \cline{1-1}
			&&&\\
			Accuracy (avg.) & 0.919 & 0.896 & 0.909\\
			Binary cross-entropy loss (avg.)& \textbf{0.165} & 0.314 &0.658\\
			\hline
			&&&\\
			 $k=5$&&&\\ \cline{1-1}
			&&&\\
			Accuracy (avg.) & \textbf{0.951} & 0.911 & 0.923\\
			Binary cross-entropy loss (avg.)& 0.218 & 0.288 &0.454\\
			\hline
			&&&\\
		    $k=8$&&&\\ \cline{1-1}
			&&&\\
			Accuracy (avg.) & 0.913 & 0.858 & 0.884\\
			Binary cross-entropy loss (avg.)& 0.168 & 0.476 &0.389\\
			\hline
	\end{tabular}
	\label{tab:cv_acc_loss}
\end{table}

The governing performance of the LSTM-based EWS is proven to be robust in the cross validation. Given different $k$ values, LSTM invariably produces the greatest accuracy and lowest loss in comparison to the baselines. In particular, EWS with LSTM achieves the best test-set accuracy of $95.1\%$ in the $5$-fold validation. And even with $3$-fold validation, LSTM obtains an accuracy of $91.9\%$ and loss of $16.5\%$ in the test set. 

\subsubsection{Back-testing}
In the back-testing, a simple trading strategy is adopted to the SSEC stock index with the aim to verify the effectiveness of the proposed EWS from a practical perspective. Assuming symmetric information between the market and the investors with a fair level of risk aversion, a market portfolio of SSEC index is constructed and held until the EWS alerts crises, and repurchased as the EWS suggests tranquility. Table \ref{tab:backtest} summarizes the expectation and standard deviation of returns together with Sharp ratios in the full sample and test set. In the absence of early warning mechanisms, the market portfolio yields expected returns of $2.3\%$ and $-0.5\%$ and standard deviations $1.48$ and $1.156\%$ in the full sample and test set, respectively. The corresponding Sharp ratios are $1.6\%$ and $-0.4\%$. By exiting the market position with respect to early warned turbulence, the strategy significantly reduces the systematic risk (indicated by the $\sigma$), which naturally results in a higher level of Sharp ratio, regardless of the predictive model. 

More importantly, back-testing once more verifies that the LSTM-based EWS outperforms the baselines and holds the greatest effectiveness and stability. Specifically, the effectiveness of LSTM is proven by its dominating Sharp ratios which improve the market portfolio by $3.8\%$ and $2.4\%$ in the full sample and test set, respectively. Meanwhile, its stability is suggested by the monotonous positive impact on the market portfolio regarding to the three portfolio measures in the risk-return horizon. Albeit the moderate improvements achieved by BPNN (Sharp ratios $4.6\%$ and $0.2\%$ in the full sample and test set) and SVR (Sharp ratios $-0.1\%$ and $0.5\%$), the two models exhibit limitations due to their weaker and fluctuating results.  

\begin{table}[h!]
	\caption{Back-testing in the full sample and test set. $E[R_{p}]$ is the expected return rate, $\sigma_{p}$ is the standard deviation and $SharpeRatio$ is given by $SharpeRatio=\frac{E[R_{p}]-R_{f}}{\sigma_{p}}$, where $R_{f}$ denotes the risk free interest rate and is set to zero in our study.}	
	\setlength{\tabcolsep}{17pt}
	\centering
	\small
	\begin{tabular}{lccc}
		\hline
		&&&\\
		&
		$E[R_{p}]$ &
		$\sigma_{p}$ & 
		$SharpeRatio$
		\\ \cline{2-4}
		full-sample&&&\\ \cline{1-1}
		&&&\\
		market portfolio & 0.023 & 1.480 & 0.016\\
		&&&\\
		EWS-LSTM &0.039&0.718&\textbf{0.054}\\
		EWS-BPNN &0.045&0.983&0.046\\
		EWS-SVR  &-0.001&0.687&-0.001\\			
		\hline
		 &&&\\
		test set&&\\ \cline{1-1}
		&&&\\
		market portfolio & -0.005 & 1.156 & -0.004\\
		&&&\\
		EWS-LSTM &0.012&0.610&\textbf{0.020}\\
		EWS-BPNN&0.004&0.625&0.002\\
		EWS-SVR&0.003&0.594&0.005\\		
		\hline
	\end{tabular}
	\label{tab:backtest}
\end{table}

\section{Conclusions}\label{conclusion}

In this study, a novel EWS with a dynamic architecture integrating the SWARCH model, two-peak thresholding and LSTM is developed to identify and predict stock market turbulence. According to the models' performance on the ten-year sample of Shanghai Stock Exchange Composite index, the following concluding remarks are emerged. 

\begin{enumerate}
	\item As one of the most powerful models handling sequential data, LSTM remains its outstanding position in the daily prediction task of stock crises. To be specific, the reliability of LSTM in this study is not only reflected by the high accuracy of $96.6\%$ and on average $2.4$ days of forewarned period, but also its stability of outperforming the baselines throughout the evaluation process in the test-set, cross-validation as well as back-testing. 
	\item In addition to a high-performing predictive model, a precise and robust crisis identification mechanism also plays the central role in facilitating the effectiveness and reliability of an EWS. By adopting the two-peak method to determine crisis cutoffs, the proposed EWS suggests a constructive alternative to current existing approaches, and yields promising crisis classifications in the Chinese stock market in comparison to the classic indicator function based on CMAX. 
	\item Stock market turbulence described by the SWARCH volatility regimes is proven to be a good crisis indicator in both theory and practice, as the proposed EWS depicts all the recorded major stock crises in the sample with significantly improved back-testing results than the market portfolio. 

\end{enumerate}

For future study, we plan to further investigate the proposed EWS structure in terms of other crisis thresholding and prediction mechanisms. At the same time, we are interested in applying the integrated EWS to predict other types of financial crises, e.g. currency or banking crises, in different frequency domains. 

\section*{Acknowledgement}
We acknowledge the support by 2016 Jiangsu Science and Technology Programme: Young Scholar Programme (No. BK20160391).

\bibliography{elsews}

\end{document}